\documentclass[11pt,draftclsnofoot,onecolumn]{IEEEtran}
\IEEEoverridecommandlockouts

\usepackage[cmex10]{amsmath}
\usepackage{amssymb}
\usepackage{stackrel}

\usepackage{graphicx}
\usepackage{cite}
\usepackage[top=0.75in,bottom=1in,left=0.625in,right=0.625in]{geometry}

\newtheorem{theorem}{Theorem}
\newtheorem{lemma}{Lemma}
\newtheorem{remark}{Remark}
\newtheorem{corollary}{Corollary}

\newcommand{\E}[1]{\mathbb{E} \left [ #1 \right ]}
\newcommand{\Et}[1]{\mathbb{E}^2 \left [ #1 \right ]}
\newcommand{\Ef}[1]{\mathbb{E}^4 \left [ #1 \right ]}

\newcommand{\si}[1]{\sum_{i=1}^{n_r}{#1}}

\newcommand{\sk}[1]{\sum_{k=1}^{n_r}{#1}}

\newcommand{\sqn}[1]{\sum_{q\neq k}{#1}}
\newcommand{\ueq}[1]{\stackrel{\text{{#1}}}{=}}
\newcommand{\ul}[1]{\stackrel{\text{{#1}}}{\leq}}

\newcommand{\ull}[1]{\stackrel{\text{{#1}}}{<}}

\newcommand{\pr}[1]{\Pr \left \{ #1 \right \} }
\newcommand{\er}{\epsilon'}
\newcommand{\err}[1]{\frac{\epsilon'}{#1}}
\newcommand{\dl}[1]{(d_{#1}^l)^{-\gamma/2}}
\newcommand{\dlt}[1]{(d_{#1}^l)^{-\gamma}}
\newcommand{\de}[1]{(d_{#1}^e)^{-\gamma/2}}

\begin{document}
%
\title{Secrecy Capacity in Large Cooperative Networks in Presence of Eavesdroppers with Unknown Locations}

%
%

\author
{
Authors\\
Sharif University of Technology, Tehran, Iran
Email: \{Authors\}@ee.sharif.edu}

\author{Amir Hossein Hadavi, Narges Kazempour, Mahtab~Mirmohseni and~Mohammad Reza Aref\\

Information Systems and Security Lab (ISSL)\\

Department of Electrical Engineering, Sharif University of Technology, Tehran, Iran\\

Email: \{ah{\_}hadavi,narges{\_}kazempour\}@ee.sharif.edu,\{mirmohseni,aref\}@sharif.edu

\thanks{}}

\maketitle

\begin{abstract}
In this paper, an extended large wireless network under the secrecy constraint is considered. In contrast to works which use idealized assumptions, a more realistic network situation with unknown eavesdroppers locations is investigated: the legitimate users only know their own Channel State Information (CSI), not the eavesdroppers CSI. Also, the network is analyzed by taking in to account the effects of both fading and path loss. Under these assumptions, a power efficient cooperative scheme, named \emph{stochastic virtual beamforming}, is proposed. Applying this scheme, an unbounded secure rate with any desired outage level is achieved, provided that the density of the legitimate users tends to infinity. In addition, by tending the legitimate users density to the infinity, the tolerable density of eavesdroppers will become unbounded too.
\end{abstract}




%
\IEEEpeerreviewmaketitle


\section{Introduction}

%
%
%
%

Nowadays, secrecy is an essential quality of service which is harder to meet in wireless networks, because their broadcast nature increases the possibility of eavesdropping. Common methods rely on using algorithms with high computational complexity that are hard to break for an adversary \cite{bloch2011physical}. Another field which focuses on the attackers with unlimited computational power is information-theoretic physical layer secrecy.
Wiretap channel, the basic model for information-theoretic secrecy, was introduced by Wyner in \cite{wyner1975wire} through which reliable and secure transmission is possible if the channel between the transmitter and the eavesdropper is the degraded version of the direct channel, i.e., between the transmitter and the receiver.

There are many research works on wireless networks with few nodes \cite{mukherjee2014principles}, but wireless systems are getting larger and larger and their exact performance analysis is getting complex, actually impossible. This leads the research community to turn into the scaling laws and analyzing the asymptotic behavior. Large wireless networks was first investigated in \cite{GupKum00} by Gupta and Kumar from the scaling laws point of view. They considered an ad hoc large network with $n$ randomly located nodes and the total rate that they achieved is $O(\sqrt{n})$.
Effects of secrecy on large wireless networks was investigated in \cite{koyluoglu2012secrecy} for the first time, where a large wireless network has been investigated that the distributions of the legitimate and eavesdropper nodes are according to the Poisson point processes with densities $\lambda_l$ and $\lambda_e$, respectively. The result of \cite{koyluoglu2012secrecy} is that the secure communication with total rate of $O(\sqrt{n})$ is possible, as long as $\lambda_e/\lambda_l=O((\log{n})^2)$, where $n$ is the number of the legitimate nodes. These works showed that it is possible to achieve the total rate that scales like $\sqrt{n}$ under per node power constraint, with and without secrecy. However, their main limiting assumption was considering a point-to-point multihopping communication which excludes the possibility of cooperation using relays.

Authors in \cite{XieKum04} proposed a cooperative scheme to achieve a total rate with near linear scaling under per node power constraint in a large wireless network without secrecy constraints. In addition, they showed the possibility of zero cost communication, i.e., unbounded total rate for fixed total power constraint.
In \cite{mirmohseni2014scaling} using active cooperative relaying based schemes and with a bound on the number of the eavesdroppers, the authors showed that zero cost secure communication is also possible. Recent developments in wireless technology (e.g., self interference cancellation, power allocation scheme at the PHY layer, proper MAC protocol for the efficient implementation of the full-duplex transmission mode \cite{zhang2015full}) support the relaying based cooperative models, in contrast to the traditional multi-hop interference limited networks.

In the model of \cite{koyluoglu2012secrecy,mirmohseni2014scaling}, the Channel State Information (CSI) is known to the legitimate transmitter. However, knowing CSI leads to the knowledge of the location of the passive eavesdroppers; that is not reasonable in many practical cases. So the natural questions here are that if zero cost secure communication is possible under unknown CSI. And, how should the cooperative strategies change to achieve this result? In addition, another important aspect of wireless network, ignored in many works, is fading. How fading affects the secrecy rate in wireless systems is a challenging question.

The secrecy rate in large networks with \emph{unknown CSI} is investigated in some recent works. In \cite{goeckel2011artificial,sheikholeslami2012physical}, the total rate of order 1 was achieved in a large wireless network with fading when CSI is not known. The authors in \cite{ccapar2012secret} took the advantage of path diversity to achieve the total rate of order $\sqrt{\frac{n}{\ln(n)}}$ in the case of unknown CSI, by limiting the number of the eavesdroppers that can be tolerated. Adding network coding has improved this result in \cite{ccapar2012network} to a scheme in which any number of eavesdroppers could be tolerated without any change in the total achievable rate. The unknown CSI assumption is also taken into account in other works such as \cite{vasudevan2010security,zhang2014asymptotic}.
However, to the best of our knowledge, none of the existing works uses relaying to achieve zero cost secure communication with unknown CSI and/or fading.

In this paper, we answer the above question affirmatively by proposing a scheme that achieves zero cost secure communication in a fading network and in the case of unknown CSI (including the eavesdroppers location).
We consider a network with $n_l$ legitimate nodes and $n_e$ eavesdroppers that are distributed according to the Poisson point processes with densities $\lambda_l$ and $\lambda_e$.
In contrary to the existing works, we achieve zero cost secure communication, i.e., unbounded total secrecy rate, by using cooperation and distributed beamforming.
In order to overcome the lack of CSI knowledge, we propose a new scheme called, \emph{stochastic virtual beamforming}.
In this 2-stage scheme, we benefit from the fading diversity by exploiting some relaying nodes near the transmitter. Actually, we design a decode and forward scenario to direct the majority of the power toward the receiver location. To make this possible, at the first step the transmitter sends the secure message to all the relaying nodes by using wiretap coding. The security of this transmission step is provided by using the distance advantage of the relaying nodes in comparison with the eavesdroppers. So we leverage the path-loss effect in a positive way. Then, at the second step, the relaying nodes accomplish a distributed beamforming by setting their transmission coefficients proportional to the complex conjugate of their channel gains to the receiver.



\section{Network Model and Preliminaries}\label{sec:definition}
Throughout the paper, use upper case letters are used for denoting the random variables and lower-case letters for their realizations. Also, superscripts $l$ and $e$ are used for denoting legitimate users and eavesdroppers, respectively. We note the desired secure rate and outage level by $R_S$ and $\epsilon$, respectively. Also, we define $\epsilon'$ to be equal to $\frac{\epsilon}{7}$.
Considering both path loss and fading effects, we use a common model for characterizing the power attenuation in wireless mediums as \cite{schwartz2004mobile}: $\frac{P_R}{P_T}=C\alpha^{2}{10}^{\frac{x}{10}}d^{-\gamma}$,
in which, $P_R$ is the received power; $P_T$ is the transmitted power; $C$ is a constant; $\alpha$ is the fading coefficient; $10^{\frac{x}{10}}$ denotes the shadow fading where $X\sim N(0,\sigma^2)$; $d$ is the distance between the transmitter and the receiver; and $\gamma$ is the path loss exponent which depends on the environment and normally $\gamma>=2$. $\alpha$ is assumed to have Rayleigh distribution with parameter $\mu$. For simplicity, we ignore the effect of shadow fading comparing with path loss effect (we remark that the shadowing effect is a random variable varying with location not with time). Also, because of different and stochastic paths between the transmitter and the receiver, the phase of the received signal (shown by $\theta$) is modeled by a uniform distribution on $[0,2\pi]$.
The letters $h$ and $d$ with appropriate subscripts and superscripts are used for indicating fading coefficients and distances, respectively. So, the channel gain from the $i$th legitimate user to the $j$th legitimate user and also, to the $k$th eavesdropper can be characterized by:
\begin{IEEEeqnarray}{rcl}
    G_{i,j}^l &=& h_{i,j}^l(d_{i,j}^l)^{-\gamma/2}e^{j\theta_{ij}^l} \\
    G_{i,k}^e &=& h_{i,k}^e(d_{i,k}^e)^{-\gamma/2}e^{j\theta_{ik}^e}.
\end{IEEEeqnarray}

We assume that the environment is isotropic. Hence the fading statistics is the same between every two nodes. We consider a network with $n_l$ legitimate nodes and $n_e$ eavesdroppers that are distributed according to the Poisson point processes with densities $\lambda_l$ and $\lambda_e$.
We consider the eavesdroppers as passive attackers with no collusion between them. In addition, we assume that neither the location nor the fading coefficient of any eavesdropper channel is not known to the legitimate users. We consider an extended wireless network. In order to establish consistency between the density of legitimate users ($\lambda_l$) and their total number ($n_l$,), we consider the network as a square with the side equal to $\sqrt{\frac{n_l}{\lambda_l}}$. Also, for the sake of simplicity we let the transmitter to be located at the center of the square.
The Rayleigh assumption for fading results in $\E{H^2}=2\mu$. Also, for simplicity we assume that the noise variances of all the channels, either legitimate or non-legitimate, are the same and equal to unity.	

\section{Main Results}
The main result of this paper is summarized in the following theorem. This theorem states that the zero-cost secure communication is possible by using our proposed scheme when eavesdroppers CSI is not known to the legitimate users. The rest of this section is devoted to the proof of this result, where we analyze the scheme in detail and derive six constraints for different parameters of the network. These constraints are consistent and can be selected step by step.
\begin{theorem}
In the extended network with fading and unknown eavesdroppers CSI (defined in Section~\ref{sec:definition}), under the constant power constraint and by letting the legitimate users density to be sufficiently large, any desired pair of secure rate and outage level denoted by ($R_S$,$\epsilon$) is achievable.
\end{theorem}
\begin{IEEEproof}	
We propose a scheme which achieves the desired result. Our proof has two steps: (i) In the first step, we consider the transmission from the transmitter (source) to the relaying nodes and guarantee a specific secure rate $R_S$ with high probability for this transmission. Our technique is based on defining two circles, denoted by $B_l$ and $B_e$, centered at the transmitter and radii $a_l$ and $a_e$, while $a_l<a_e$ (see Fig.\ref{fig1}). Then $\lambda_l$, $\lambda_e$, $a_l$, and $a_e$ are chosen such that the following three requirements are provided. First, with the probability greater than $1-{\epsilon'}$, no eavesdropper lies in $B_e$. Second, with the probability greater than $1-\epsilon'$ at least $n_r$ legitimate users lie in $B_l$. Third, the difference of the worst legitimate channel and the best eavesdropper channel be greater than $R_S$ with a probability greater than $1-\epsilon'$.
(ii) In the second step, we analyze the rate from the relay nodes to the receiver and guarantee the second rate using the cooperation of $n_r$ relaying nodes. Actually, we make this distributed Multiple-Input Single-Output Single-antenna Eavesdropper (MISOSE) situation to concentrate the most of the transmitted power in a neighboring region of the receiver. It can be deduced from our following calculations that by increasing $n_r$, both $R_S$ and $\epsilon$ can be improved, i.e., increased and decreased, respectively.
\subsection{Step 1: First rate analysis}
In this step, we guarantee a secure rate $R_S$ for the transmission from the transmitter to the relaying nodes with an outage level of $2\epsilon'$. To make this possible, we choose the radius of the circles $B_l$ and $B_e$ in a way that even with considering possible exacerbating effects of the fading, the difference between the capacities of the worst legitimate channel and the best eavesdropper channel be greater than $R_S$ which is done by obtaining proper upper and lower bounds on $a_l$ and $a_e$, respectively. Hence, the following constraint must hold:
\begin{equation}
\min_{1\le i\le n_r , 1\le j\le n_e}{C_i^l-C_j^e}> R_S
\end{equation}
where $C_i^l$ is the rate of the link from the transmitter to the $i$-th legitimate user and $C_j^e$ is the rate of the link from the transmitter to the $j$-th eavesdropper. To simplify the analysis, we work with a suboptimum problem and we guarantee the following two inequalities:
\begin{align}
\min_{1<i<n_r}{C^l_i}&>(1+\rho)R_S ,\label{leg_rate1} \\
\max_{1<j<n_e}{C^e_j}&<\rho R_S.		\label{eav_rate1}
\end{align}
in which, $\rho$ is an arbitrary positive constant and the problem can be optimized over $\rho$. Now, to establish \eqref{leg_rate1} and \eqref{eav_rate1}, we present appropriate upper and lower bounds on $a_l$ and $a_e$, respectively, where each bound holds with an outage level of $\epsilon'$.
\subsubsection{The legitimate rate analysis}
In the following theorem, considering the constraint on $C^l_i$, we derive an appropriate upper bound on $a_l$.
\begin{theorem}[Upper bound on $a_l$]\label{THM_al}
A sufficient condition for having \eqref{leg_rate1}, with an outage level of $\epsilon'$, is:
\begin{equation}\label{al_bound}
a_l<\Big( \frac{-P_T\mu \ln{(1-\frac{\epsilon'}{n_r})}}{2^{(1+\rho)R_S}-1} \Big)^{\frac{1}{\gamma}}.
\end{equation}
\end{theorem}
\begin{IEEEproof} Using the union bound we guarantee the outage level of $\frac{\epsilon'}{n_r}$ for the rate of each of the $n_r$ relaying nodes in $B_l$, to guarantee the outage level of $\epsilon'$ for the minimum of these rates. We write for one of them, chosen arbitrarily:
\begin{align}
\log{(1+P_Th^2 d_l^{-\gamma})}&>\log{(1+P_Th^2a_l^{-\gamma})}>(1+\rho)R_S \Rightarrow\nonumber\\
 a_l^{\gamma}&<\frac{Ph^2}{2^{(1+\rho)R_S}-1}.\label{des_al}
\end{align}
where $d_l$ and $h$ are the distance and the channel gain (respectively) between the transmitter and the chosen relay, so $d_l<a_l$.
We require the validity of \eqref{des_al} with a probability more than $1-\err{n_r}$. With respect to the Rayleigh distribution assumption for $h$, the distribution of $h^2$ is exponential with parameter $\frac{1}{\mu}$. Hence, with the probability of $1-\err{n_r}$, we have: $h^2>-\mu \ln{(1-\frac{\epsilon'}{n_r})}$.
Thus, if $a_l$ satisfies the bound in \eqref{al_bound}, the inequality \eqref{des_al} holds with a probability more than $1-\err{n_r}$.
\end{IEEEproof}
\subsubsection{Network Layering scheme and eavesdropper rate analysis}
To analyze the eavesdroppers rates, one can follow a similar approach to what presented in the previous part. However, the eavesdroppers are distributed in all around the network and their distances from the transmitter vary from $a_e$ to the radius of the network. Hence, following the same approach would yield a loose bound on $a_e$. For deriving a tighter bound we propose a network layering scheme.
In this scheme, as shown in Fig.~\ref{fig1}, the network is divided to a number of layers and the eavesdroppers rates in each layer is analyzed separately. To be precise, the $k$-th layer is defined as the region of the network between the radii $2^{k-1}a_e$ and $2^{k}a_e$. We repeat this procedure till the boundary of the network. In the following, we propose a lower bound on $a_e$ using this idea. We denote the number of layers by $K_L$ and the $k$-th layer by $L_k$.
\begin{figure}
\centering
\includegraphics[width=5cm,keepaspectratio]{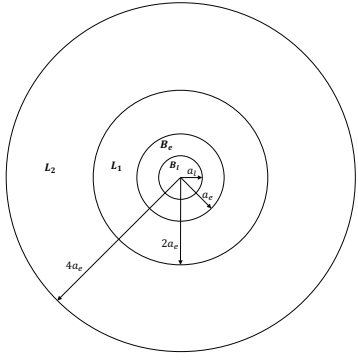}
\caption{{The inner and outer circles and the two first layers in the network layering scheme}}
\label{fig1}
\end{figure}
The area of $L_k$, denoted by $S_k$, is equal to $S_k=\pi\left(2^{2k}-2^{2(k-1)}\right)a_e^2=3\pi 2^{2(k-1)}a_e^2$.
To bound the number of eavesdroppers in each layer we present two following lemmas.

\begin{lemma}[Number of eavesdroppers in each layer]\label{lem_num1}
For any positive constant $\beta_k$, define $t_k$ as:
\begin{equation}
t_k\triangleq\big(\frac{\beta_k}{\epsilon'}.\frac{1}{\lambda_eS_k}\big)^{\frac{1}{2}} . \label{tcond}
\end{equation}
Then, with a probability larger than $1-\frac{\epsilon'}{\beta_k}$, the number of eavesdroppers in $L_k$ (denoted as $n_{e,k}$) satisfies:
\begin{equation}\label{numbound}
n_{e,k}<(1+t_k)\lambda_eS_k.
\end{equation}
\end{lemma}
\begin{IEEEproof}
Considering Poisson distribution of eavedroppers locations, we use Chebyshev's inequality for $n_{e,k}$ to write:
\begin{align*}
&\pr{n_{e,k}>\lambda_eS_k+t_k{\lambda_eS_k}}\\
&< \pr{\vert  n_{e,k}-\lambda_eS_k\vert>t_k\sqrt{\lambda_eS_k}\sqrt{\lambda_eS_k}}\\
&<\frac{1}{t_k^2\lambda_eS_k}.
\end{align*}
\end{IEEEproof}

\begin{lemma}\label{lem_num2}
If in Lemma~\ref{lem_num1}, we set $\beta_k$s such that $\sum_{k=1}^{K_L}{\frac{1}{\beta_k}}<1$ holds,
  then, the inequality \eqref{numbound} will be valid for all the layers, with a probability larger than $1-\epsilon'$.
\end{lemma}
\begin{IEEEproof}
Using the union bound for the undesired event in each layer, we can bound the global undesired event. Therefore, the probability that the inequality \eqref{numbound} does not hold in at least one layer is bounded by:
\begin{equation*}
\sum_{k=1}^{K_L}{\frac{\epsilon'}{\beta_k}}={\epsilon'}\sum_{k=1}^{K_L}{\frac{1}{\beta_k}}.
\end{equation*}
Assuming the condition introduced in the lemma on $\beta_k$s, this quantity will be less than $\epsilon'$.
\end{IEEEproof}

Now we derive a proper bound on $a_e$ using the above lemmas. In fact, each layer imposes a lower bound on $a_e$ and the largest lower bound is the main constraint on $a_e$. Our technique of deriving these bounds is summarized in the following. We divide the tolerable error $\epsilon'$ between all the layers, dedicating the tolerable error $\epsilon_k=\frac{\epsilon'}{2^k}$ to the $L_k$, and we find a proper lower bound to guarantee the outage probability of $\epsilon_k$ for this layer. In the analysis of each layer we apply the union bound for the eavesdroppers in that layer. Finally, we apply the union bound on the outage events of these layers to find a bound on the probability of the total outage event. This total outage probability is less than $\epsilon'$, because of how $\epsilon_k$s are allocated.

Using Lemmas~\ref{lem_num1} and \ref{lem_num2}, we work with $(1+t_k)\lambda_eS_k$ as the maximum number of eavesdroppers in $L_k$.
The following lemma gives the constraint on $a_e$ concluded from $L_k$.
\begin{lemma}\label{lem3}
Given the inequality \eqref{numbound}, a sufficient condition to have,
\begin{equation}\label{Lk}
\max_{j\in L_k}{C^e_j}<\rho R_S
\end{equation}
with a probability greater than $1-\epsilon_k$, is the following constraint on $a_e$:
\begin{align}\label{LK2}
a_e&>a_e^{(k)}\triangleq 2^{-(k-1)}\Big(\frac{-P_T\mu}{2^{\rho R_S}-1}\ln{(\frac{\epsilon'}{2^{k}\lambda_e(1+t_k)S_k})}\Big)^{\frac{1}{\gamma}}.
\end{align}
\end{lemma}
\begin{IEEEproof}
To guarantee the outage level of $\frac{\epsilon'}{2^k}$ for the validity of \eqref{Lk}, relying on the union bound, we guarantee the outage level of $\frac{\epsilon'}{2^k n_{e,k}}$ for the validity of the following inequality:
\begin{equation}
C^e_j<\rho R_S
\end{equation}
for each of the eavesdroppers, e.g., the $j$-th eavesdropper,  in this layer. Considering the condition \eqref{numbound} on $n_{e,k}$, it suffices to guarantee the outage level of $\frac{\epsilon'}{2^k \lambda_e(1+t_k)S_k}$ for each of the eavesdroppers. For one of the eavesdroppers, arbitrarily chosen, we write:
\begin{align*}
\log{(1+Ph^2d_e^{-\gamma})}<\log{(1+Ph^2(2^{-{k-1}}a_e)^{-\gamma})}<\rho R_S.
\end{align*}
The first inequality is deduced from $d_e<2^{k-1}a_e$, in which $d_e$ is the distance between the transmitter and the chosen eavesdropper.  For simplicity, the other indices are eliminated.
We want the second inequality to be valid with a probability greater than $1-\frac{\epsilon'}{2^k \lambda_e(1+t_k)S_k}$. With the same probability, for the coefficient $h^2$, considering its exponential distribution, the following inequality holds:
\begin{equation*}
h^2<h^2_{\text{max}}\triangleq-\mu \ln{\left(\frac{\epsilon'}{2^{k}\lambda_e(1+t_k)S_k}\right)}.
\end{equation*}
Hence, to provide the desired outage level, it suffices to guarantee the inequality
\begin{equation*}
\log{(1+Ph^2_{\text{max}}(2^{-(k-1)}a_e)^{-\gamma})}<\rho R_S,
\end{equation*}
by proper choice of $a_e$. With a little algebraic efforts and displacing the variables, the recent inequality can be converted to \eqref{LK2}.
\end{IEEEproof}
Now, we put all the results together in the following theorem. Then, from this theorem and by some substitutions and calculations, we conclude the Corollary \ref{corol} in which an appropriate lower bound is finalized for $a_e$.

\begin{theorem}\label{THM_ae}
Given the coefficients $t_k$ and $\beta_k$ consistent with the assumptions presented in Lemmas~ \ref{lem_num1} and \ref{lem_num2} and by choosing
\begin{equation}\label{unibound}
a_e>\max_{1\leq k\leq K_L}{a_e^{(k)}},
\end{equation}
the rate inequality \eqref{eav_rate1} holds with an outage probability less than $2\epsilon'$.
\end{theorem}
\begin{IEEEproof}
Note that the inequality \eqref{eav_rate1} is valid if and only if the one in \eqref{Lk} is valid. We define the binary random variable $O_k$ as
\begin{equation}
O_k = \left\{
\begin{array}{rl}
0 & \text{if } \max_{j\in L_k}{C^e_j}<\rho R_S,\\
1 & \text{if } \max_{j\in L_k}{C^e_j}>\rho R_S.
\end{array} \right.
\end{equation}
In addition, we define the binary random variable $Q_k$ as
\begin{equation}
Q_k = \left\{
\begin{array}{rl}
0 & \text{if } n_{e,k}<(1+t_k)\lambda_eS_k,\\
1 & \text{if } n_{e,k}>(1+t_k)\lambda_eS_k.
\end{array} \right.
\end{equation}
Now, using the union bound, we write:
\begin{equation}\label{sum}
\pr{\max_{1<j<n_e}{C^e_j}>\rho R_S}<\sum_{k=1}^{L_K}{\pr{O_k=1}}.
\end{equation}
Also, we expand the occurrence probability of $O_k$ in $Q_k$ as the following:
\begin{align}
\pr{O_k=1}&=\sum_{i=0,1}{\pr{Q_k=i}\pr{O_k=1|Q_k=i}}\nonumber\\
&\ul{(a)}\pr{O_k=1|Q_k=0}+\pr{Q_k=1}.\label{expand}
\end{align}
Considering that the value of the probability function is not never more than the unity, the terms $\pr{Q_k=0}$ and $\pr{O_k=1|Q_k=1}$ in the inequality (a) are replaced by 1. First, we investigate the first term of \eqref{expand}. Given \eqref{unibound}, the offered  sufficient condition presented in the Lemma \ref{lem3} (\eqref{LK2}) holds for all the layers.	So, considering this lemma, we have:
\begin{equation}
\pr{O_k|Q_k=0}<\frac{\epsilon'}{2^k}.
\end{equation}
By summing up in all the layers, we write:
\begin{equation}\label{t1}
\sum_{k=1}^{L_K}{\pr{O_k|Q_k=0}}<\sum_{k=1}^{L_k}{\frac{\epsilon'}{2^k}}<{\epsilon'}.
\end{equation}
Now we consider the second term in \eqref{expand}. For the summation of these terms,  according to the Lemma \ref{lem_num2}, we have:
\begin{equation}\label{t2}
\sum_{k=1}^{K_L}{\pr{Q_k=1}}<{\epsilon'}.
\end{equation}
Now the proof is completed by applying \eqref{expand} in \eqref{sum} and then using the inequalities \eqref{t1} and \eqref{t2}.
\end{IEEEproof}

\begin{corollary}\label{corol}
By choosing
\begin{equation}\label{final_ae}
a_e = \frac{(P_T\mu)^{\frac{1}{\gamma}}}{(2^{\rho R_S}-1)^\frac{1}{\gamma}} (\ln{\frac{-\ln{(1-\epsilon')^6}}{\epsilon'}}+\sqrt{\frac{2\epsilon'}{\left(-\ln{(1-\epsilon')^3}\right)^3}} ),
\end{equation}
the outage probability of \eqref{eav_rate1} is less than $2\epsilon'$.
\end{corollary}	
\begin{IEEEproof}
We start from the right side of the constraint \eqref{unibound} in the recent theorem and insert the value of $a_e^{(k)}$ from \eqref{LK2}. Also, we replace $t_k$ by its calculated value from \eqref{tcond} and for any $1\le k\le K_L$, we set:
\begin{equation}
\beta_k = 2^{k}.
\end{equation}
It's clear that by this choice, the required condition in the Lemma~\ref{lem_num2} for $\beta_k$s is established. Furthermore, we replace $\lambda_e$ by its value from \eqref{final_lambe}. Now, we can write:
\begin{align*}
a_e &= \max_{1\le k \le K_L}{a_e^{(k)}}\\
&=\left( \frac{P\mu}{2^{\rho R_S}-1} \right)^{\frac{1}{\gamma}}\max_{1\le k \le K_L}{-2^{-(k-1)}\ln{\frac{\epsilon'}{2^k \lambda_e S_k(1+t_k)}}}\\
&=\left( \frac{P\mu}{2^{\rho R_S}-1} \right)^{\frac{1}{\gamma}}\max_{1\le k \le K_L}{2^{-(k-1)}\times\ln{\frac{-3\times 2^{3k-2}\ln{(1-\epsilon')}+\left( \frac{4}{-3\epsilon'\ln{(1-\epsilon')}} \right)^{\frac{1}{2}}\beta_k^{\frac{1}{2}}}{\epsilon'}}}\\
&\ueq{(a)} \left( \frac{P\mu}{2^{\rho R_S}-1} \right)^{\frac{1}{\gamma}}\max_{1\le k \le K_L}{2^{-(k-1)}\ln{\left( c_1 2^{3k}+c_2 2^{\frac{k}{2}}\right)}}\\
&= \left( \frac{P\mu}{2^{\rho R_S}-1} \right)^{\frac{1}{\gamma}}\max_{1\le k \le K_L}{2^{-(k-1)}\ln{c_1 2^{3k}\left( 1+\frac{c_2}{c_1}2^{\frac{-5}{2}k} \right)}}\\
&\ull{(b)} \left( \frac{P\mu}{2^{\rho R_S}-1} \right)^{\frac{1}{\gamma}}\max_{1\le k \le K_L}{2^{-(k-1)}(3k\ln{2}+\ln{c_1}+\frac{c_2}{c_1}2^{-\frac{5}{2}k})}\\
&\ueq{(c)}\left( \frac{P\mu}{2^{\rho R_S}-1} \right)^{\frac{1}{\gamma}}\left(3\ln{2}+\ln{c1}+\frac{c_2}{4\sqrt{2}c_1}\right)
\end{align*}
In the equation (a), the following definitions are used:
\begin{align}
c_1 &\triangleq \frac{-3\ln{(1-\epsilon')}}{4\epsilon'},\\
c_2 &\triangleq \sqrt{\frac{4}{-3\epsilon'\ln{(1-\epsilon')}}} .
\end{align}
Inequality (b) is deduced from $\ln{x}<x-1$
for $x>0$ and $x\neq1$. The inequality (c) is obtained by considering that the argument of maximization is a decreasing function in $k$, so it takes its maximum at $k=1$. Substituting this value for $k$, the last line is obtained.
\end{IEEEproof}
By replacing $c_1$ and $c_2$ by their values we reach to the same value in the relation
\eqref{final_ae}.  Now we set $a_e$ equal to this value. Hence, the desired condition in the theorem \ref{THM_ae} will be valid, too.

\begin{remark}
We remark that the proposed idea for layering the network is really effective. In fact, as it's clear from \eqref{final_ae}, the final constraint on $a_e$ is independent of the number of the eavesdroppers (and so from the size of the network). Therefore, by extending the network and so increasing the total number of the eavesdroppers, it's not necessary to limit the eavesdropper-free region anymore.
\end{remark}

\subsection{Step 2: Second rate analysis}
In this subsection we analyze the second rate, i.e., forward rate, and from it we derive the proper constraint on $n_r$.
\subsubsection{Fading calculations}
In this part we pick arbitrarily one of the eavesdroppers and do the calculations for it. We denote the channel gain vector between the source and the legitimate and the non-legitimate users, respectively, as:
 \begin{align}
\underline{g}^l &=\left(\dl{1}h_1^le^{j\theta_1^l},\dots,\dl{n_r}lh_{n_r}^le^{j\theta_{n_r}^l}\right), \\
\underline{g}^e &=(\de{1}h_1^ee^{j\theta_1^e},\dots,\de{n_r}h_{n_r}^ee^{j\theta_{n_r}^e}).
\end{align}
We assume that the distance between every two users is greater than half of the wavelength. This assumption yields the uncorrelation of different fading gains and phases \cite{goldsmith2005wireless}. We establish a virtual and distributed MISOSE situation using adequate relaying nodes. This scheme has two advantage comparing with conventional Multiple-Input Multiple-Output (MIMO) schemes. First, it does not need the devices to be equipped with multiple antennas. Second, as noted in \cite{bloch2011physical}, the maximum number of antennas in practical MIMO systems has physical limitation. But, in this scheme we can exploit more relaying nodes and benefit more from channel diversity. In \cite{shafiee2007achievable}, for MISOSE situation with ergodic capacity criterion and known CSI and only for legitimate users, it has been proved that the efficient strategy for the beamforming of the transmission vector is to align it in the direction of the fading vector. Thus, by a similar technique, we align the beamforming vector in the direction of complex conjugate of channel gain vector, i.e., $(\underline{g}^l)^*$, to maximize the correlation between these two vectors. In order to control the total consumed power, we set the beamforming vector equal to $\frac{(\underline{g}^l)^*}{n_r}$. Given only legitimate users CSI, it is a reasonable strategy.

\textbf{Message transmission scheme}:
For the message set $M=[1:2^{nR}]$ and for any $m\in M$, a proper codeword $X^n$ generated from Wyner wiretap coding is chosen and transmitted by the transmitter. We denote the average power of the transmitter by $P_T$. The relaying nodes decode their received sequence to obtain the transmitted message $m$. In the next step, the $i$-th relaying node uses the same codebook to send the sequence $F_i=\frac{1}{\sqrt{n_r}}\dl{i}h_iX^ne^{-j\theta_i^l}$
in $n$ transmission intervals.
So the power consumed by the $i$-th relaying node and the total consumed power equal to:
\begin{align}
 P_i&=\frac{1}{n_tn_r}\sum_{t=1}^{n_t}{\dlt{i}h_i^2|X(t)|^2}=\frac{\dlt{i}h_i^2 }{n_r}P_T,\\
 P_T^{\text{(tot)}}&=\sum_{i=1}^{n_r}{P_i}=(\frac{1}{n_r}\sum_{i=1}^{n_r}{\dlt{i}h_i^2})P_T.
 \end{align}
 Furthermore, the received signals at the end of the $t$-th transmission interval are:
 \begin{align*}
	 Y(t)&=(\frac{1}{\sqrt{n_r}}\sum_{i=1}^{n_r}{\dlt{i}(h_i^l)^2})X(t) ,\\
	 Z(t)&=(\frac{1}{\sqrt{n_r}}\sum_{i=1}^{n_r}{\dl{i}\de{i} h_i^lh_i^ee^{j(\theta_i^e-\theta_i^l)}})X(t).
\end{align*}
Finally, the received powers at the legitimate user and the eavesdroppers are:
\begin{align}
P_l &=(\frac{1}{\sqrt{n_r}}\sum_{i=1}^{n_r}{\dlt{i}h_i^2})^2P_T,  \label{Pld} \\
P_e &=\Big\vert\frac{1}{\sqrt{n_r}}\sum_{i=1}^{n_r}{\dl{i}\de{i}h_i^lh_i^ee^{j({\theta_i^e-\theta_i^l})}}\Big\vert^2P_T . \label{Ped}
\end{align}
Now we consider these two recent random variables (i.e., $P_l,P_e$) and give bounds on their expected values and variances. Having these in hand, we can use a bounding inequality, like Chebyshev's inequality, to predict the behavior of these two quantities with high probability.

\textbf{Probabilistic results}:
Based on the assumptions we noted previously about the fading coefficients, we proved the following bounds for the expected value and variances of $P_l$ and $P_e$. The proof is provided in  appendix \ref{app1}.
\begin{theorem}\label{THM1}
By appropriate choices for $\eta$ and $\nu$, the following bounds hold:
\begin{align}
\frac{\E{P_l}}{P_T}&>\eta n_r(d_{TR}+a_l)^{-2\gamma},\label{g1}\\
\frac{\E{P_e}}{P_T}&<{\eta}(a_e-a_l)^{-\gamma}(d_{TR}-a_l)^{-\gamma},\label{g2}\\
\frac{\sigma^2(P_l)}{P_T^2}&<{\nu^2}{n_r}(d_{TR}-a_l)^{-4\gamma},\label{g3}\\
\begin{split}\label{g4}
\frac{\sigma^2(P_e)}{P_T^2}&<{\nu^2}(a_e-a_l)^{-2\gamma}(d_{TR}-a_l)^{-2\gamma}.
\end{split}
\end{align}
where, $d_{TR}$ is the distance between the transmitter and the receiver. It is assumed that the receiver is out of the inner circle ($B_l$).
\end{theorem}
To continue, we look for the sufficient number of relaying nodes in order to attain the secure rate $R_S$ with outage probability less than $2\epsilon'$. We use Chebyshev's inequality to establish proper bounds on probability of the undesired events defined on the amount of $P_l$ and $P_e$. We wish to have $C_l-\max_{i\in \mathcal{E}}{C_i^e}>R_S$ with a probability greater than $1-\epsilon'$.
But, for the sake of simplicity, we guarantee the following bounds, each with the probability of $1-\epsilon'$:
\begin{align}
C_l&>(1+\kappa)R_S,\label{leg_rate2} \\
\max_{i\in \mathcal{E}}{C_i^e}&<\kappa R_S.\label{unleg_rate2}
\end{align}
where, $\kappa$ is an arbitrary positive constant which can be optimized if necessary. Now, we derive the proper bounds on the network parameters by analyzing the above limitations.
Instead of \eqref{unleg_rate2}, using a union bound approach, we consider the following constraint for each eavesdropper:
\begin{equation}\label{eav_sec_rate}
 C_i^e<\kappa R_S
 \end{equation}
with the probability of $1-\err{n_e}$.

\subsubsection{Legitimate rate analysis}
The constraint in \eqref{leg_rate2} implies $\pr{C_l=\log{(1+P_l)}<(1+\kappa)R_S}<\er$,
or equivalently:
\begin{equation}
\pr{P_l<2^{(1+\kappa)R_S}-1}<\er.\label{legdes}
\end{equation}
Now, we apply Chebyshev's inequality and drive a lower bound on $n_r$ which guarantees \eqref{legdes}.
Noting the expected value and the variance of $P_l$ by $\eta_l$ and $\nu_l^2$, respectively, we apply the inequalities of Theorem \ref{THM1} for these two values. First, we write:
\begin{align*}
\pr{P_l<2^{(1+\kappa)R_S}-1}&\ull{(a)}\pr{P_l<\eta_l-\alpha\nu_l}\\
&<\pr{|P_l-\eta_l|>\alpha\nu_l}\ull{(b)}\frac{1}{\alpha^2}\ul{(c)}\er.
\end{align*}
where, (b) follows from Chebyshev's inequality,
for (c) we set: $\alpha=\sqrt{\frac{1}{\epsilon'}}$, and
for (a), it's sufficient to have: $2^{(1+\kappa)R_S}-1<\eta_l-\alpha\nu_l$,
or equivalently: $\nu_l<\frac{1}{\alpha}(\eta_l-2^{(1+\kappa)R_S}+1$.
Considering \eqref{g1} and \eqref{g3}, it's sufficient to establish the following chain:
\begin{align*}
\nu_l &\ull{(d)} {\nu}{\sqrt{n_r}}(d_{TR}-a_l)^{-2\gamma}P_T\\
&\ull{(e)} \frac{1}{\alpha}(\eta{n_r}(d_{TR}+a_l)^{-2\gamma}P_T-2^{(1+\kappa)R_S}+1)\\
&\ull{(f)}\frac{1}{\alpha}(\eta_l-2^{(1+\kappa)R_S}+1).
\end{align*}
where, (d) and (f) are deduced from Theorem \ref{THM1}. We establish (e) by choosing $n_r$ sufficiently large. After some algebraic calculations, (e) can be written as the following quadratic inequality in $\sqrt{n_r}$:
\begin{align*}
&n_r(\eta(d_{TR}+a_l)^{-2\gamma}P_T) - \sqrt{n_r}(\alpha{\nu}(d_{TR}-a_l)^{-2\gamma}P_T) \\
&\qquad\qquad\qquad\qquad\qquad\qquad\qquad\quad -2^{(1+\kappa)R_S}+1 > 0.
\end{align*}
in which only one of the two roots is positive and so acceptable. By choosing $n_r$ greater than the square of this root, we reach a constraint on $n_r$ presented in the following theorem.
\begin{theorem}[Lower bound for $n_r$]\label{THM2}
A sufficient condition on $n_r$ for guaranteeing \eqref{leg_rate2} with an outage level of $\epsilon'$ is to have:
\begin{align}
n_r &> \frac{(d_{TR}-a_l)^{-4\gamma}}{4\eta^2(d_{TR}+a_l)^{-4\gamma}}({\frac{\nu}{\sqrt{\epsilon'}} + \sqrt{\zeta}} )^2,\label{nrlegcond}\\
\zeta &= \frac{\nu^2}{\epsilon'}+4\eta\frac{(d_{TR}+a_l)^{-2\gamma}}{P_T(d_{TR}-a_l)^{-4\gamma}}(2^{(1+\kappa)R_S}-1).\nonumber
\end{align}
\end{theorem}
In order to get an intuition from the behavior of this constraint, we put a simplifying assumption on $a_l$, which makes this constraint independent of $a_l$. For this, we assume:
\begin{equation}\label{al_assum}
a_l<d_{TR}/2.
\end{equation}
We justify this assumption by noting that if the receiver lies in $B_l$, it is not necessary to exploit the \emph{stochastic virtual beamforming} scheme. Actually in this situation, based on the calculations for the first rate, the message can be delivered securely to the receiver by direct transmission.
Here, for the sake of simplicity, after choosing a valid value for $a_l$, i.e., a value which satisfies the constraint in \eqref{al_bound}, we divide it by two. By this choice, we can send the secure message directly to the receiver whenever the receiver is in the distance of at most $2a_l$. Therefore, we use our proposed scheme, i.e., the \emph{stochastic virtual beamforming}, only when \eqref{al_assum} holds. Using this, constraint \eqref{nrlegcond} is turned to the following simplified version. The proof is simple and is completed by bounding $d_{TR}\pm a_l$ properly.
\begin{corollary}
A simplified sufficient condition on $n_r$ to guarantee \eqref{leg_rate2} with the outage level of $\epsilon'$ is to have:
\begin{equation}\label{nrlegcond_simp}
n_r > \frac{81}{4\eta^2}\Big(\frac{\nu}{\sqrt{\epsilon'}} + \sqrt{\frac{\nu^2}{\epsilon'}+\frac{4\eta}{P_T}d_{TR}^{2\gamma}(2^{(1+\kappa)R_S}-1)} \Big)^2.
\end{equation}
\end{corollary}

\subsubsection{Eavesdropper rate analysis}
Now, we proceed in a similar way to obtain another constraint to guarantee \eqref{eav_sec_rate} for the eavesdropper rate with high probability. As noted previously, for the arbitrarily chosen eavesdropper, we wish to have: $\pr{\log(1+P_e)>\kappa R_S}=\pr{P_e>2^{\kappa R_S}-1}<\err{n_e}$.
Similar to the previous part, we use $\eta_e$ and $\nu_e^2$ to denote the expected value and the variance of $P_e$. We start with:
\begin{align*}
\pr{P_e>2^{\kappa R_S}-1}&\ull{(a)}\pr{P_e>\eta_e+\alpha\nu_e}\\
&<\pr{|P_e-\eta_e|>\alpha\nu_e}\ull{(b)}\frac{1}{\alpha^2}\ul{(c)}\err{n_e},
\end{align*}
where (b) follows from Chebyshev's inequality and for (c) we set $\alpha=\sqrt{\frac{n_e}{\epsilon'}}$.
Similar to the previous part, for establishing (a), it's sufficient to have $\nu_e<\frac{1}{\alpha}(2^{\kappa R_S}-\eta_e-1)$.
Now, considering \eqref{g2} and \eqref{g4}, it is sufficient to establish the following chain,
\begin{align*}
\nu_e &\ull{(d)} \nu(a_e-a_l)^{-\gamma}(d_{TR}-a_l)^{-\gamma}P_T\\
&\ull{(e)}\frac{1}{\alpha}\left( 2^{\kappa R_S}-\eta(a_e-a_l)^{-\gamma}(d_{TR}-a_l)^{-\gamma}P_T-1 \right)\\
&\ull{(f)}\frac{1}{\alpha}\left( 2^{\kappa R_S}-\eta_e-1 \right).
\end{align*}
By some substitution and assuming the other parameters to be constant, the inequality (e) can be converted to a constraint on $n_e$, as follows:
\begin{align}\label{final_ne}
n_e < {\epsilon'}\big( \frac{2^{\kappa R_s}-\eta(a_e-a_l)^{-\gamma}(d_{TR}-a_l)^{-\gamma}-1}{\nu(a_e-a_l)^{-\gamma}(d_{TR}-a_l)^{-\gamma}P_T} \big)^2.
\end{align}

\subsection{Poisson calculations}
\subsubsection{Constraint related to the inner circle} We must have at least $n_r$ legitimate relaying nodes in the circle $B_l$, where $n_r$ is chosen appropriately regarding the former constraint in \eqref{nrlegcond}.
In the following, we start by bounding the probability of undesirable event, i.e., having less than $n_r$ nodes in $B_l$, using Chebyshev's inequality. Then, using this bound, we derive a sufficient condition for $\lambda_l$, in order to keep the probability of undesirable event less than ${\epsilon'}$. By defining the $k_l$ as the number of legitimate nodes in $B_l$, we write:
\begin{align}
\pr{k_l<n_r}&<\pr{\vert k_l-\lambda_l\pi a_l^2 \vert > \lambda_l\pi a_l^2-n_r}\nonumber\\
&=\pr{\vert k_l-\lambda_l\pi a_l^2 \vert > \sqrt{\lambda_l\pi a_l^2}\frac{\lambda_l\pi a_l^2-n_r}{\sqrt{\lambda_l\pi a_l^2}}}\nonumber\\
&<\frac{\lambda_l\pi a_l^2}{(\lambda_l\pi a_l^2-n_r)^2}.\label{undes}
\end{align}

To satisfy the outage probability constraint, it suffices to set \eqref{undes} less than or equal to ${\epsilon'}$. In the equality case, we reach the following equation from which a lower bound on $\lambda_l$ is deduced. By satisfying this constraint, we have at least $n_r$ legitimate nodes in the inner circle, $B_l$, with probability larger than $1-{\epsilon'}$.
\begin{align}
&\lambda_l^2(\pi^2a_l^4)-\lambda_l(2n_r+\frac{1}{\epsilon'})\pi a_l^2+n_r^2=0 \Rightarrow\nonumber\\
&\lambda_l=\frac{n_r+\frac{1}{2\epsilon'}\pm\sqrt{\left(n_r+\frac{1}{2\epsilon'}\right)^2-n_r^2}}{\pi a_l^2}\label{sol}
\end{align}
Note that the smaller solution in \eqref{sol} is not acceptable, because it yields values for $\lambda_l$ which are lower than $\frac{n_r}{\pi a_l^2}$. So we work with the greater solution. As a sufficient condition, we can choose $\lambda_l$ to be greater than this solution. Therefore,
\begin{IEEEeqnarray}{l}
\lambda_l > \frac{n_r+\frac{1}{2\epsilon'}+\sqrt{(n_r+\frac{1}{2\epsilon'})^2-n_r^2}}{\pi a_l^2}=
\nonumber\\
\frac{n_r}{\pi a_l^2}\big(\underbrace{1+\frac{1}{2\epsilon' n_r}+\sqrt{(1+\frac{1}{2\epsilon' n_r})^2-1}}_{\beta_l(\epsilon)}\big).
\end{IEEEeqnarray}
By the above definition of $\beta_l(\epsilon)$, we summarize the constraint as:
\begin{IEEEeqnarray}{l}\label{final_lambl}
\lambda_l>\beta_l(\epsilon).\frac{n_r}{\pi a_l^2}.
\end{IEEEeqnarray}

\subsubsection{Constraint related to the outer circle} As mentioned previously we need the circle $C^e$ to be free of eavesdroppers, with probability larger than $1-{\epsilon'}$. By defining $k_e$ as the number of eavesdroppers in $B_e$, we want to have:
\begin{IEEEeqnarray}{l}
\pr{k_e=0}>1-{\epsilon'} \Rightarrow e^{-\lambda_e\pi a_e^2}>1-{\epsilon'}
\end{IEEEeqnarray}
which results in the constraint:
\begin{IEEEeqnarray}{l}\label{final_lambe}
\lambda_e<\frac{-\ln(1-{\epsilon'})}{\pi a_e^2}
\end{IEEEeqnarray}
The main six constraints for $a_l$, $a_e$, $n_r$, $n_e$, $\lambda_l$ and $\lambda_e$ are given in \eqref{al_bound}, \eqref{final_ae}, \eqref{nrlegcond_simp}, \eqref{final_ne}, \eqref{final_lambl} and \eqref{final_lambe}, respectively. To achieve any desired pair of $(R_S,\epsilon)$, we proceed as follows. First, we set $n_r$ satisfying \eqref{nrlegcond_simp}, which just depends on $R_S$ and $\epsilon$ and not the other five parameters. Knowing $n_r$, we choose $a_l$ properly from its constraint in \eqref{al_bound}. Then, the required $\lambda_l$ is calculated by inserting the value of $n_r$ and $a_l$ in \eqref{final_lambl}. In addition, minimum of $a_e$ is computed by knowing $R_S$ and $\epsilon$ from \eqref{final_ae}. Then, the maximum tolerable amount of $\lambda_e$ is derived from \eqref{final_lambe}. It is seen that by tending $\lambda_l$ to infinity, the maximum tolerable density of eavesdroppers tends to infinity, too. This completes the proof.
\end{IEEEproof}
\appendix[proof of theorem \ref{THM1}] \label{app1}
First, we consider the problem without the path loss effect and analyze the four desired quantities for this case. Then, we add the effect of path loss and update the previous bounds for this case.

The following fading vectors are in fact the simplified versions of channel gain vectors, $\underline{g}^l$ and $\underline{g}^e$, when all the coefficients related to path loss effect are substituted by unity:
\begin{align}
\underline{h}^l=(h_1^le^{j\theta_1^l},\dots,h_{n_r}^le^{j\theta_{n_r}^l}),\\
\underline{h}^e=(h_1^ee^{j\theta_1^e},\dots,h_{n_r}^ee^{j\theta_{n_r}^e}).
\end{align}
In the following lemma, the values of the four desired quantities are given when the path loss effect is eliminated.
\begin{lemma}
Under the assumptions noted in the paper for the fading coefficients, the mean and the variance of $P_l$ and $P_e$ satisfy the following constraints, when the path loss effect is eliminated:
\begin{align}
   \frac{\E{P_l}}{P_T}&=(n_r-1)\Et{H^2}+\E{H^4}\nonumber\\
   &>(n_r-1)\Et{H^2}+\Et{H^2}
   =n_r\Et{H^2}, \\
   \frac{\E{P_e}}{P_T}&=\Et{H^2},\\
   \frac{\sigma^2(P_l)}{P_T^2}&=4n_r\left(\E{H^4}\Et{H^2}-\Ef{H^2} \right)+\dots = \text{{O}}\left(n_r\right),\label{Varl}\\
   \frac{\sigma^2(P_e)}{P_T^2}&=\frac{1}{n_r}\Ef{H^2} +\dots = \text{{O}}\left(1\right) . \label{Vare}
\end{align}
\end{lemma}
\begin{IEEEproof}
Analysis for mean of $P_l$:
\begin{align*}
\frac{\E{P_l}}{P_T}&=\E{\left(\frac{1}{\sqrt{n_r}}\sum_{i=1}^{n_r}{(h_i^l)^2}\right)^2} \\
&= \frac{1}{n_r}\E{\sum_{i=1}^{n_r}\sum_{k=1}^{n_r}{(h_i^l)^2(h_k^l)^2}}\\
&=\frac{1}{n_r}\left(n_r\E{H^4}+n_r(n_r-1)\Et{H^2}\right)\\
&=(n_r-1)\Et{H^2}+\E{H^4}\\
&>(n_r-1)\Et{H^2}+\Et{H^2}=n_r\Et{H^2}.
\end{align*}

Analysis for mean of $P_e$:
\begin{align*}
\begin{split}
\frac{\E{P_e}}{P_T}&=\E{\left\vert\frac{1}{\sqrt{n_r}}\sum_{i=1}^{n_r}{h_i^lh_i^ee^{j({\theta_i^e-\theta_i^l})}}\right\vert^2}\\
&=\frac{1}{n_r}\E{\sum_{i=1}^{n_r}\sum_{k=1}^{n_r}{h_i^lh_i^eh_k^lh_k^ee^{j(\theta_i^e-\theta_i^l-\theta_k^e+\theta_k^l)}}}\\
&\stackrel{\text{{(a)}}}{=}\frac{1}{n_r}\sum_{i=1}^{n_r}{\E{(h_i^l)^2}\E{(h_i^e)^2}}+\frac{n_r(n_r-1)}{n_r}\sum_{i=1}^{n_r}\sum_{k\neq i}{\E{h_i^lh_i^eh_k^lh_k^e}\underbrace{\E{e^{j\theta_i^e}}}_{0}\E{e^{j(-\theta_i^l-\theta_k^e+\theta_k^l)}}}\\
&=\frac{n_r}{n_r}\Et{H^2}=\Et{H^2}.
\end{split}
\end{align*}
where (a) is deduced from the uniform distribution assumption for the random phases and the independence of the fading coefficients and also the phases of the legitimate and non-legitimate channels.

Analysis for variance of $P_l$:
\begin{align*}
\frac{\sigma^2(P_l)}{P_T^2}&=\E{\left( \frac{\left(\sum_{i=1}^{n_r}{(h_i^l)^2}\right)^2}{n_r}-\left(\E{H^4}+(n_r-1)\Et{H^2}\right) \right)^2}\\
&=\frac{1}{n_r^2}\E{\left(\sum_{i=1}^{n_r}{h_i^4}+\sum_{k=1}^{n_r}\sum_{q\neq k}{h_k^2h_q^2}
-n_r\E{H^4}-n_r(n_r-1)\Et{H^2}\right)^2}\\
&=\frac{1}{n_r^2}\E{\left(\underbrace{\sum_{i=1}^{n_r}{h_i^4-\E{H^4}}}_{S_1}+\underbrace{\sk{\sqn{h_k^2h_q^2-\Et{H^2}}}}_{S_2}\right)^2}\\
&=\frac{1}{n_r^2}\left( \E{S_1^2}+2\E{S_1S_2}+\E{S_2^2} \right),
\end{align*}
where
\begin{align*}
2\E{S_1S_2}&\ueq{}4n_r(n_r-1)\E{(H_1^4-\E{H^4})(H_1^2H_2^2-\Et{H^2})}\\
&=4n_r(n_r-1)(\E{H^6}\E{H^2}-\E{H^4}\Et{H^2}),\\
\E{S_1^2}&\stackrel{\text{{}}}{=}n_r\E{(H^4-\E{H^4})^2}\\
&=n_r(\E{H^8}-\Et{H^4}),\\
\E{S_2^2}&\ueq{} n_r(n_r-1)(\Et{H^4}-\Ef{H^2})+4n_r(n_r-1)(n_r-2)\E{(H_1^2H_2^2-\Et{H^2})(H_1^2H_3^2-\Et{H^2})}\\
&= n_r(n_r-1)(\Et{H^4}-\Ef{H^2})+4n_r(n_r-1)(n_r-2)\left( \E{H^4}\Et{H^2}-\Ef{H^2} \right).
\end{align*}
Hence:
\begin{align*}
\frac{\sigma^2(P_l)}{P_T^2}&=\frac{1}{n_r^2}\Big( n_r(\E{H^8}-\Et{H^4})+n_r(n_r-1)(\Et{H^4}-\Ef{H^2})\\
&\quad +4n_r(n_r-1)(n_r-2)\left(\E{H^4}\Et{H^2}-\Ef{H^2}\right)\\
&\quad +4n_r(n_r-1)(\E{H^6}\E{H^2}-\E{H^4}\Et{H^2})\Big)\\
&\ueq{}\frac{4(n_r-1)(n_r-2)}{n_r}\Big(\E{H^4}\Et{H^2}-\Ef{H^2}\Big)+\underbrace{\cdots}_\text{} .
\end{align*}

Analysis for variance of $P_e$:
\begin{align*}
\frac{\sigma^2(P_e)}{P_T^2}&=\E{\left(\left\vert\frac{1}{\sqrt{n_r}}\sum_{i=1}^{n_r}{h_i^lh_i^ee^{j({\theta_i^e-\theta_i^l})}}\right\vert^2-\Et{H^2}\right)^2}\\
&=\frac{1}{n_r^2}\E{\left(\underbrace{\si{(h_i^l)^2(h_i^e)^2-\Et{H^2}}}_{A_1}+\underbrace{\sk{\sqn{h_k^lh_k^eh_q^lh_q^ee^{j(\theta_k^l-\theta_k^e-\theta_q^l+\theta_q^e)}}}}_{A_2}\right)^2}\\
&=\frac{1}{n_r^2}\left( \E{A_1^2}+2\E{A_1A_2}+\E{A_2^2} \right),
\end{align*}
where
\begin{align*}
\E{A_1^2}&\ueq{}n_r\E{\left(H_1^2H_2^2-\Et{H^2}\right)^2}=n_r\left(\Et{H^4}-\Ef{H^2}\right),\\
\E{A_2^2}&\ueq{}n_r(n_r-1)\Ef{H^2},\\
2\E{A_1A_2}&\ueq{}0.
\end{align*}
results in:
\begin{align*}
\frac{\sigma^2(P_e)}{P_T^2}&=\frac{n_r-1}{n_r}\Ef{H^2}+\frac{1}{n_r}\left( \Et{H^4}-\Ef{H^2} \right)\ueq{}\Ef{H^2}+\dots .
\end{align*}
\end{IEEEproof}
\begin{corollary}\label{corolfade}
There are positive coefficients $\eta$ and $\nu$, such that for the without path loss case, we have:
\begin{align}
\frac{\E{P_l}}{P_T}&>\eta n_r, \\
\frac{\E{P_e}}{P_T}&=\eta,\\
\frac{\sigma^2(P_l)}{P_T^2}&<{\nu^2}{n_r},\\
\frac{\sigma^2(P_e)}{P_T^2}&<\nu^2.
\end{align}
\end{corollary}
\begin{IEEEproof}
For $\eta$ we just set
\begin{equation}
\eta=\Et{H^2}=4\mu^2.
\end{equation}
The existence of $\nu$ is proved by considering \eqref{Varl} and \eqref{Vare} and the final expression is obtained for these two quantities and the finiteness of the Rayleigh distribution moments. By substituting the required moments in that final expressions, this coefficient can be chosen and it can be shown that it is not too large. Note that these coefficients just depend on the statistical behavior of Rayleigh distribution and its parameter and are chosen independent of other parameters of our scheme like $R_S$, $\epsilon$ and $P_T$.
\end{IEEEproof}

Now, we prove the main results proposed in the Theorem \ref{THM1}, i.e., the results for the complete model when the path loss effect is taken in to account. First, we prove \eqref{g1} and \eqref{g2}. The proofs of \eqref{g3} and \eqref{g4} are more elaborate and needs two lemmas to be proved.
\begin{IEEEproof} [Proof of \eqref{g1} and \eqref{g2}]
Considering the geometry of the network, we have the following common bounds for all $d_i^l$s and $d_i^e$s:
\begin{align}
 d_{TR}-a_l<&d_i^l<d_{TR}+a_l,\label{bou1}\\
 a_e-a_l<&d_i^e.\label{bou2}
 \end{align}

 By using these bounds, we extract the quantities related to the path loss effect from the summations in the expressions of \eqref{Pld} and \eqref{Ped}, so that the remaining terms in the summations change to the same expressions related to the case without considering the path loss effect. Now, using the results stated in Corollary \ref{corolfade} and by considering the linearity and monotonicity of the expected value function, \eqref{g1} and \eqref{g2} are simply concluded.
\end{IEEEproof}
Now we prove the two variance results (\eqref{g3} and \eqref{g4}). First, we present the following lemmas.
\begin{lemma}\label{lem_rayl}
For any non-negative random variable $H$, with positive mean, the following inequality is true:
 \begin{equation}
 \E{H^3} \geq \E{H^2}\E{H}.\label{raylst}
 \end{equation}
 \end{lemma}
\begin{IEEEproof}
Since $H\geq 0$, by using Cauchy-Schwartz inequality for the two random variables ${H^\frac{1}{2}}$ and  $H^{\frac{3}{2}}$, we can write:
\begin{align*}
\E{H^3}\E{H}&=\E{\left(H^{3/2}\right)^2)}\E{\left(H^{1/2}\right)^2}\\
&\geq \Et{H^2}
\geq \E{H^2}\Et{H}.
\end{align*}
Now, considering its positivity, we divide the above relations by $\E{H}$ to obtain the inequality \eqref{raylst}.
\end{IEEEproof}

\begin{lemma}\label{lem_var}
For every two i.i.d. random variables $X$ and $Y$ with positive mean and variance and for any two positive constants $a$ and $b$ such that $a<b$, the following inequality is true:
\begin{equation}\label{varineq}
\text{Var}[(aX+bY)^2]<b^4\text{Var}[(X+Y)^2].
\end{equation}
\end{lemma}
\begin{IEEEproof}
By expanding the left side, we show that substituting $a$ by $b$ will increase the variance.
\begin{align*}
\text{Var}[(aX+bY)^2]=&\text{Var}[a^2X^2+b^2Y^2+2abXY]\\
=&\E{\left( a^2(X^2-\E{X^2})+b^2(Y^2-\E{Y^2})+2ab(XY-\E{XY}) \right)^2}\\
=&\E{a^4(X^2-\E{X^2})^2}+\E{b^4(Y^2-\E{Y^2})^2} +\E{4a^2b^2(XY-\E{XY})^2}\\
& + 2a^2b^2\E{(X^2-\E{X^2})(Y^2-\E{Y^2})} + 4a^3b\E{(X^2-\E{X^2})(XY-\E{XY})}\\
& + 4ab^3\E{(Y^2-\E{Y^2})(XY-\E{XY})}.
\end{align*}
In the last equality, the three first terms are clearly non-negative and increasing the coefficients, will increase the total result. So, substituting $a$ by $b$ increases the Variance. According to the independence assumption, the forth term can be decomposed to two expected value terms, which both of them are zero. The fifth and the sixth sentence have a similar form. We show below that the fifth term is always positive. A similar argument is true for the sixth term.
\begin{align*}
\E{(X^2-\E{X^2})(XY-\E{XY})}&\ueq{(a)}\E{X^3}\E{Y}-\E{X^2}\E{X}\E{Y}\\
&=\E{Y}\left(\E{X^3}-\E{X^2}\E{X}\right)\stackrel{\text{(b)}}{>}0.
\end{align*}
where (a) is deduced from the independence assumption; (b) is concluded from Lemma \ref{lem_rayl} and the positivity of the mean of $Y$. So, the fifth and also the sixth sentence are positive and therefore all the six sentences have a non-negative value. Hence, replacing $a$ by $b$ will increase the amount of the variance and the validity of \eqref{varineq} is established.
\end{IEEEproof}
\begin{IEEEproof} [proof of \eqref{g3} and \eqref{g4}]
Using induction, the recent lemma can be generalized to any number of random variables. Considering the independence of the fading coefficients and their Rayleigh distribution and using the inequalities \eqref{bou1} and \eqref{bou2}, the generalization of Lemma \ref{lem_var} results in \eqref{g3} and \eqref{g4}.	
\end{IEEEproof}

\bibliographystyle{./IEEEtran}
\bibliography{./IEEEabrv,./iwcit16}

%

%
%
%




\end{document}